\documentclass[twocolumn,english,pra,showpacs,superscriptaddress,longbibliography]{revtex4-1}
\usepackage{amsfonts}
\usepackage{amsmath}
\usepackage{graphics}
\usepackage{amssymb}
\usepackage[colorlinks, citecolor=blue,linkcolor=red]{hyperref}
\usepackage{color}
\usepackage{hyperref}
\usepackage{textcomp}
\usepackage[rightcaption]{sidecap}
\usepackage{subfigure}
\usepackage{float,csquotes,mathtools}
\usepackage[rightcaption]{sidecap}
\usepackage{graphicx}
\usepackage{lipsum}

\begin{document}
\title{$\mathcal{PT}$-assisted control of Goos-Hänchen shift in cavity magnomechanics}
%PT-assisted control of the Goos-Hänchen shift
\author{Shah Fahad}
\affiliation{Department of Physics, Zhejiang Normal University, Jinhua, Zhejiang 321004, China}
\author{Gao Xianlong}
\email {gaoxl@zjnu.edu.cn}\affiliation{Department of Physics, Zhejiang Normal University, Jinhua, Zhejiang 321004, China}

%%%%%%%%%%%%%%%%%%%%%%%%%%%%%%%%%%%%%%%%%%%%%%%%%%%%%%%%%%%%%%%%%%%%%%%%%%%%%%%%%%%%%%%%%%%%%%%%%%%%%%%%%%%%%%%%%%%%%%%%%%%
\setlength{\parskip}{0pt}
\setlength{\belowcaptionskip}{-10pt}
\begin{abstract}
We propose a scheme to manipulate the Goos-H\"{a}nchen shift (GHS) of a reflected probe field in a non-Hermitian cavity magnomechanical system. The platform consists of a yttrium-iron-garnet sphere coupled to a microwave cavity, where a strong microwave drive pumps the magnon mode and a weak field probes the cavity. The traveling field's interaction with the magnon induces gain, yielding non-Hermitian dynamics. When the traveling field is oriented at $\pi/2$ relative to the cavity's $x$-axis, the system realizes $\mathcal{PT}$ symmetry; eigenvalue analysis reveals a third-order exceptional point ($\mathrm{EP}_3$) at a tunable effective magnon-photon coupling. Under balanced gain-loss and finite effective magnomechanical coupling, we demonstrate coherent control of the GHS by steering the system across the $\mathcal{PT}$-symmetric transition and through $\mathrm{EP}_3$ via the effective magnon-photon coupling, enabling pronounced enhancement or suppression of the lateral shift. Furthermore, we show that without effective magnomechanical coupling, the system exhibits a second-order exceptional point ($\mathrm{EP}_2$) with a distinct GHS phase transition. This phase transition vanishes when the effective magnomechanical coupling exceeds a parametric threshold, where strong absorption at resonance suppresses the GHS. We also identify the intracavity length as an additional control parameter for precise shift tuning. Notably, the $\mathcal{PT}$-symmetric configuration yields substantially larger GHS than its Hermitian counterpart. These results advance non-Hermitian magnomechanics and open a route to GHS-based microwave components for quantum switching and precision sensing.

\end{abstract}
%\pacs{42.50.Pq, 42.50.Gy, 67.85.Hj, 71.70.Ej}
\date{\today}
\maketitle
\section{Introduction}
Magnons, the quantized spin excitations in magnetic materials~\cite{Serga_2010}, have emerged as a key platform for investigating macroscopic quantum phenomena~\cite{LENK_2011, Liu_2019, CUi_2019}. By uniting magnons, photons, and phonons, cavity magnomechanics serves as a versatile testbed bridging quantum information science, magnonics, cavity quantum electrodynamics, and quantum optical studies~\cite{zuo_2024}. This integrated framework enables the exploration of both semi-classical and quantum dynamics, leading to diverse applications such as ultrahigh-sensitivity sensing~\cite{Colombano-2020PRL, Zhang2024}, generation of squeezed and entangled states~\cite{Fan-2023, Hussain-2022, Li_2022, Li-EntPRL, Li_2019_sqeez}, ultraslow light~\cite{Lu-2023, Cui-2019-SloW}, and the transfer, storage, and retrieval of quantum states~\cite{Sarma_2021, Qi-2021}. Recent experiments have demonstrated fundamental phenomena such as magnomechanically induced absorption and transparency~\cite{Zhang_2016_cavity}, along with advanced functionalities including mechanical bistability~\cite{Shen-2022PRL}, magnonic frequency combs~\cite{Xu-2023PRL}, and efficient microwave–optical transduction~\cite{Shen-2022PRL}.

Non-Hermitian physics has emerged as a vibrant research field that explores systems governed by non-Hermitian Hamiltonians $H\neq H^\dagger$, whose complex eigenvalues reflect energy exchange with the environment~\cite{Ozdemir2019, Feng2017}. Mostafazadeh introduced pseudo-Hermiticity, wherein a non-Hermitian Hamiltonian $H$ with a discrete spectrum satisfies $H^\dagger = \eta H \eta^{-1}$ for a linear Hermitian operator $\eta$, guaranteeing that its eigenvalues are either real or appear in complex-conjugate pairs~\cite{Mostafazadeh-I, Mostafazadeh-II}. $\mathcal{PT}$-symmetric Hamiltonians, characterized by $[H, \mathcal{PT}] = 0$, constitute a significant subset of pseudo-Hermitian systems~\cite{Konotop-RMP, HOEP_PRA_2021, Zhang2025} that can exhibit entirely real energy spectra under balanced gain and loss~\cite{Bender-1999, Bender-PRL}. Varying a control parameter drives the system through a second-order exceptional point ($\mathrm{EP}_2$), triggering spontaneous $\mathcal{PT}$-symmetry breaking through the coalescence of eigenvalues and their corresponding eigenvectors. This bifurcation demarcates the transition from the unbroken $\mathcal{PT}$-symmetric phase (real spectrum) to the broken $\mathcal{PT}$ phase (complex-conjugate spectrum)~\cite{Bender-2013, Liu_2017, Midya_2021}. $\mathrm{EP}_2$s have been rigorously investigated across diverse physical systems, revealing phenomena such as topological energy transfer~\cite{Xu2016}, loss-induced lasing revival~\cite{Peng-2014}, enhanced sensitivity~\cite{Chen2017}, coherent absorption~\cite{Sweeney-2019, Changqing-2021}, single-mode lasing~\cite{Liang-2014, Hossein-2014}, and optimized energy harvesting~\cite{Fernandez-Alcazar2021}.
Beyond $\mathrm{EP}_2$, higher-order exceptional points ($\mathrm{EP}_n$, $n>2$) exhibit stronger non-Hermitian degeneracies where $n$ eigenstates coalesce~\cite{Heiss_2016, HOEP_PRA_2021, Zhang2025, Zhang_PRB_2019, Demange_2012, Qian-2025}. These higher-order EPs enable greater sensitivity~\cite{Hodaei2017, Zeng_2021}, enhanced spontaneous emission~\cite{Lin_PRL}, rapid entanglement generation~\cite{Li-2023PRL} and richer topological features~\cite{Ding-PRX, Delplace_PRL}. Recent advances demonstrate sophisticated control over phenomena such as microwave field transmission using engineered non-Hermitian systems~\cite{Wang_2023, CUi_2019}. In this work, we investigate the physics of $\mathrm{EP}_3$ on Goos–H\"{a}nchen shift (GHS) as a distinct and measurable signature of the optical response in a $\mathcal{PT}$-symmetric non-Hermitian cavity magnomechanical (CMM) system.

The GHS is an optical phenomenon governing the lateral shift (parallel to the plane of incidence) of light beams at interfaces between media structures~\cite{Shui_2019, Zia_2015_PRA, Asiri_2023, Zia_2010_Coherent}. This shift originates from spatial dispersion in transverse magnetic (TM) or transverse electric (TE) reflection coefficients and interference among angular spectrum components (i.e., phase variations in reflection coefficients)~\cite{Wang_2008, Asiri_2023, Rafi_2024}. The GHS was first theoretically predicted by Picht~\cite{Picht-1929}, experimentally verified by Goos and Hänchen~\cite{Goos-1947, Goos-1949}, and later rigorously formalized by Artmann via the stationary-phase method~\cite{Artmann-1948}. The GHS magnitude’s sign (positive/negative) emerges from the system’s parameter-dependent phase response~\cite{Li-2003PRL, Liu-2006}. The GHS has been extensively explored and manipulated in numerous quantum systems, including quantum dots~\cite{Idress-PRA}, atom-cavity quantum electrodynamics (QED) setups~\cite{Zia_2010_Coherent}, and cavity optomechanical platforms~\cite{Muhib-PRA, Anwar-PRA}. Its applications span multiple domains, such as quantum sensing~\cite{Le-PRL}, neutron optics~\cite{Haan-PRL, McKay-2025}, temperature measurement~\cite{Lu_2022}, and humidity sensing~\cite{Wang-2016}. More recently, the GHS has been investigated within a Hermitian CMM system~\cite{waseem_Goos_2024}. However, the evolution of the GHS in $\mathcal{PT}$-symmetric CMM systems—particularly across their phase transitions and at a third-order exceptional point ($\mathrm{EP}_3$)—remains uncharacterized. Here, we analyze a hybrid three-mode system featuring the coherent interplay of photon–magnon and magnon–phonon couplings. Under experimentally feasible parameters, we demonstrate that these interactions govern the lateral shift, yielding a phase-selective response. The GHS exhibits pronounced enhancement or suppression at the $\mathrm{EP}_3$ and across the $\mathcal{PT}$ phases, governed by the system's non-Hermitian eigenvalue structure. Our results establish coherent control over the GHS in a non-Hermitian CMM system, presenting a novel paradigm for tunable microwave devices and enhanced sensing.

We consider a non-Hermitian CMM system that spans the broken $\mathcal{PT}$ phase, $\mathrm{EP}_3$, and the unbroken $\mathcal{PT}$ phase. The system comprises a microwave cavity with two fixed mirrors and an embedded yttrium iron garnet (YIG) sphere. A $z$-axis bias magnetic field $B_z$ excites the magnon modes in the YIG sphere, coupling to cavity photons through magnetic dipole interactions, while magnetostrictive interactions mediate the magnon-phonon coupling via lattice deformation. Non-Hermiticity is introduced through a traveling field driving the magnon mode, inducing gain. Under gain-loss balance conditions and finite effective magnomechanical coupling, an $\mathrm{EP}_3$ emerges when the effective magnon-photon coupling strength is tuned to a critical value.

Using the stationary-phase method, we then systematically analyze the GHS in the reflected probe field across the different $\mathcal{PT}$-symmetric phases. Our results reveal strong phase-dependent behavior: the GHS is significantly enhanced in the unbroken $\mathcal{PT}$ phase compared to the broken phase, consistent with the previous report of enhanced beam shifts in atomic-ensemble-based $\mathcal{PT}$-symmetric cavity~\cite{Zia_2015_PRA}. In contrast, the $\mathrm{EP}_3$ regime exhibits a suppressed GHS. We further demonstrate that the GHS can be actively controlled across $\mathcal{PT}$ phases by tuning the intracavity length. Finally, a direct comparison under identical parameters shows that the GHS in the $\mathcal{PT}$-symmetric system exceeds that of the corresponding Hermitian configuration~\cite{waseem_Goos_2024}, highlighting the enhanced phase sensitivity enabled by the gain–loss balance.

The rest of the manuscript is structured as follows: Section II introduces the system Hamiltonian. We derive the optical susceptibility using the semiclassical Heisenberg–Langevin equations and then employ the stationary-phase method to calculate the GHS of the reflected probe field. Sec. III presents the results and discussion, focusing on the output field spectra and the control of the GHS via the system’s parameters. Sec. IV concludes.
\section{Theoretical Model}
\subsection{System Hamiltonian}
\begin{figure}[tp]
\includegraphics[width=\linewidth]{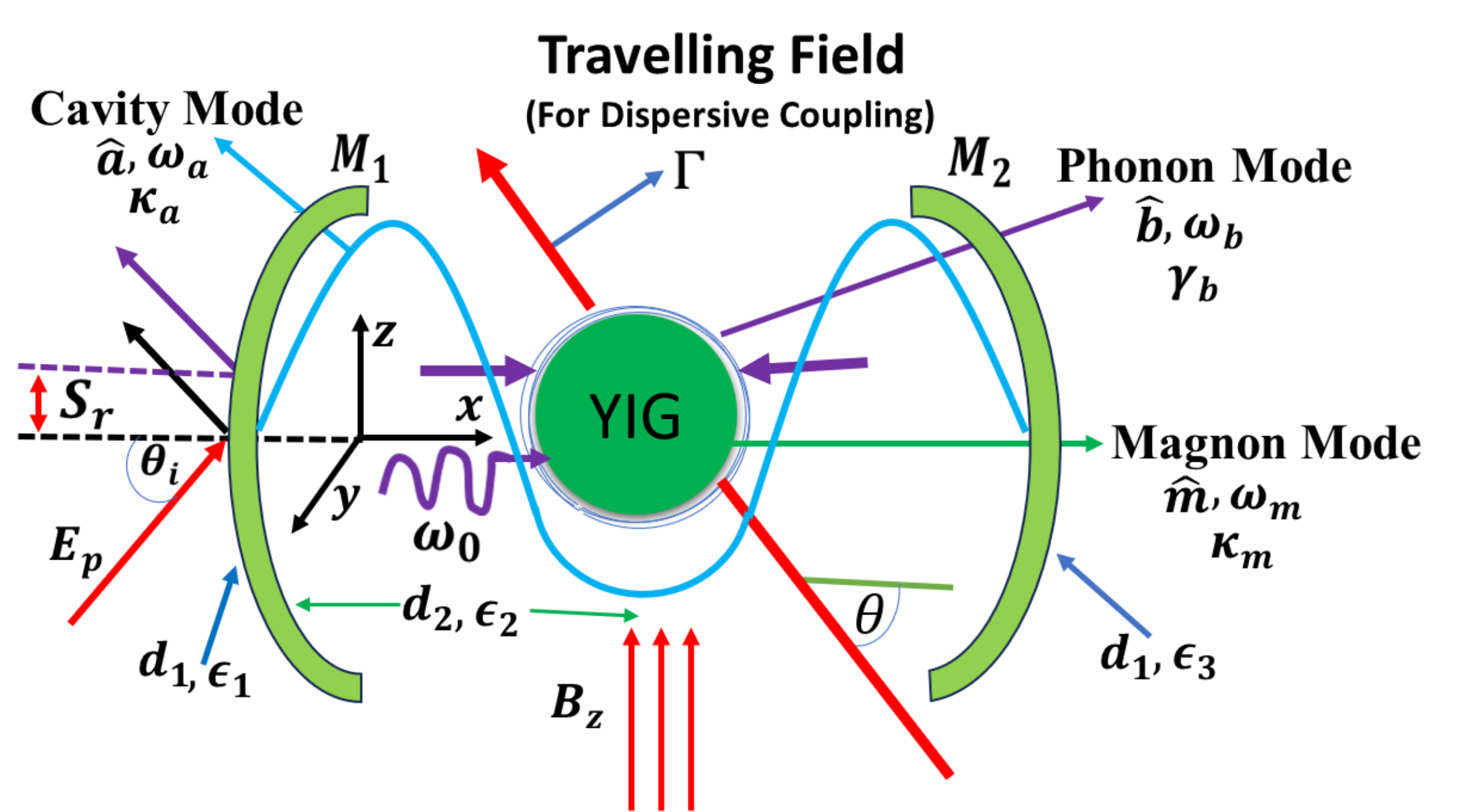}
\caption{Schematic illustration of a non-Hermitian CMM system featuring an embedded YIG sphere. The cavity mode ($\hat{a}$ with frequency $\omega_a$ and dissipation rate $\kappa_a$) is subjected to a bias magnetic field $B_z$ along the $z$-axis, which excites the magnon mode ($\hat{m}$ with frequency $\omega_m$ and gain $\kappa_m$). Magnon-photon coupling arises from magnetic dipole interactions. Magnetostrictive interactions induce the phonon mode ($\hat{b}$ with frequency $\omega_b$ and dissipation rate $\gamma_b$), facilitating magnon-phonon coupling that is further enhanced by an $x$-axis microwave drive field (frequency $\omega_0$). The perpendicular magnetic fields—cavity ($B_y$), drive ($B_x$), and bias ($B_z$)—are depicted. Non-Hermiticity emerges from a traveling field characterized by incident angle $\theta$ and coupling strength $\Gamma$. A TE-polarized probe field $E_p$ is incident on mirror $M_1$ at angle $\theta_i$. The lateral displacement experienced by the reflected probe field during total internal reflection constitutes the GHS, denoted by $S_{r}$.}
\label{fig1}
\end{figure}
We consider a hybrid non-Hermitian CMM system comprising a single-mode cavity ($\hat{a}$, frequency $\omega_{a}$) with an embedded YIG sphere, as shown in Fig.~\ref{fig1}. The YIG sphere is positioned at an antinode of the cavity's microwave magnetic field to optimize magnon-photon coupling. The setup includes two nonmagnetic mirrors ($M_1$ and $M_2$) separated by a fixed distance $d_2$. $M_2$ is perfectly reflective, while $M_1$ is partially reflective. Both mirrors have identical thickness $d_1$ and permittivities $\epsilon_1$ and $\epsilon_3$, respectively, while the intracavity medium is characterized by an effective permittivity 
$\epsilon_2$. This three-layer configuration is analogous to cavity optomechanics~\cite{Muhib-PRA} and atomic systems~\cite{Wang_2008}. A uniform bias magnetic field $B_z$ applied along the $z$-axis excites magnon modes ($\hat{m}$, frequency $\omega_m$) in the YIG sphere. These magnon modes couple with the cavity photon modes via magnetic dipole interaction. The magnetization dynamics induced by the magnon excitation lead to lattice deformation through magnetostriction, giving rise to phonon modes ($\hat{b}$, frequency $\omega_b$). Consequently, magnon-phonon coupling arises from the magnetostrictive interaction. The magnon mode is driven by microwave field (frequency $\omega_0$, amplitude $\eta = \sqrt{5N}\gamma B_{0}/4$, with external magnetic field $B_0$, total spins $N$, and gyromagnetic ratio $\gamma$). A weak probe field, with amplitude  $E_{p} = \sqrt{2P\kappa_{a}/\hbar\omega_{p}}$ (corresponding to power $P$, frequency $\omega_p$, and incident angle $\theta_i$), drives the cavity mode. The non-Hermitian behavior of the system arises from the direct interaction with a traveling field, which introduces gain in the magnon mode (with mass $m_{m}$ and frequency $\omega_{m}$), characterized by a coupling strength $\Gamma = \frac{\omega_{a}}{d_{2}}\sqrt{\hbar/\omega_{m} m_{m}}$) ~\cite{pramanik2020, Ge-PRA-2013, Teklu-20218}.
 
Under the rotating-wave approximation, the total Hamiltonian in a frame rotating at the drive frequency $\omega_{0}$ becomes~\cite{fahad2025}:
\begin{align}
\hat{\mathcal{H}} &= \hbar\Delta_{a}\hat{a}^{\dagger}\hat{a}+ \hbar\Delta_{m}\hat{m}^{\dagger}\hat{m}+ \hbar\omega_b\hat{b}^{\dagger}\hat{b} \nonumber\\ 
& +\hbar(g_{ma}-i \Gamma \mathrm{e}^{i \theta})(\hat{a}\hat{m}^{\dagger}+\hat{a}^{\dagger}\hat{m}) + \hbar g_{mb}\hat{m}^{\dagger}\hat{m}(\hat{b}^{\dagger}+\hat{b}) \nonumber\\ 
&+ i \hbar\eta(\hat{m}^{\dagger}-\hat{m}) + i\hbar E_{p}(\hat{a}^{\dagger}\mathrm{e}^{-i \Delta_{p} t}-\hat{a}\mathrm{e}^{i \Delta_{p} t}).\label{simplified-H}
\end{align}
Here $\Delta_{a} = \omega_{a} - \omega_{0}, \quad \Delta_{m} = \omega_{m} - \omega_{0},\ \text{and} \quad \Delta_{p} =\omega_{p}-\omega_{0}$. $\hat{a}$, $\hat{b}$, $\hat{m}$ ($\hat{a}^\dagger$, $\hat{b}^\dagger$, $\hat{m}^\dagger$) are annihilation (creation) operators; $g_{ma}$ ($g_{mb}$) is the magnon-photon (magnon-phonon) coupling.

We derive the semiclassical Heisenberg-Langevin equations for the operators $\hat{O} \in \{\hat{a},\hat{m},\hat{b}\}$ to describe the average cavity response, neglecting both thermal and quantum noise:
\begin{equation}
\frac{d\hat{O}}{dt}
= \frac{i}{\hbar}[\hat{\mathcal{H}}, \hat{O}]- \beta\hat{O},
\label{GenericForm}
\end{equation}
where $\beta$ represents the gain ($\beta<0$) or decay rate ($\beta>0$). The canonical commutation relation $[\hat{O}, \hat{O}^{\dagger}] = 1$ holds for $\hat{O} \in \{\hat{a},\hat{m},\hat{b}\}$. Replacing the operators with their expectation values, $O(t) \equiv \langle \hat{O}(t) \rangle$ ($O = a, m, b$)~\cite{Xiong2015}, yields:
 \begin{align}
\dot{a}&= -(i\Delta_{a}+\kappa_{a})a-(ig_{ma}+\Gamma \mathrm{e}^{{i}\theta })m + E_{p}\mathrm{e}^{-i\Delta_{p} t},\nonumber\\ 
\dot{m}&= -(i\Delta_{m} - \kappa_{m})m -(ig_{ma}+\Gamma \mathrm{e}^{i\theta})a-ig_{mb}m(b^\ast + b) + \eta,\nonumber\\ 
\dot{b}&= -(i\omega_{b}+\gamma_{b})b -ig_{mb}m^{\ast}m.\label{HLEs-2} 
\end{align}
Here $\kappa_a$ ($\gamma_b$) is the photon (phonon) dissipation rate, and $\kappa_m$ the magnon gain. The steady-state solutions are:
\begin{align}
a_{s}&= \frac{-(ig_{ma} +\Gamma {e}^{i\theta})m_{s}}{( i\Delta_{a}+\kappa_{a})},\nonumber\\
m_{s}& = \frac{-(ig_{ma}+\Gamma \mathrm{e}^{i\theta}){{a}_{s}} +\eta}{(i\Delta_{s} - \kappa_{m})},\nonumber\\
b_{s} &= \frac{-ig_{mb} |{m_{s}}|^2}{(i\omega_{b}+\gamma_{b})},\label{steady states}
\end{align}
where $\Delta_{s}= \Delta_{m} + g_{mb}({b}^{\ast}_{s} +{b_s})$ is the effective magnon-phonon detuning. After linearizing Eq.~(\ref {HLEs-2}) and retaining the first-order terms, we obtain:
 \begin{align}
\delta\dot{a}& = -(i\Delta_{a} + \kappa_{a})\delta{a}-(ig_{ma} + \Gamma {e}^{i\theta})\delta{m} + E_{p}\mathrm{e}^{-i \Delta_{p} t},\nonumber\\
\delta\dot{m}&= - (i\Delta_{s} - \kappa_{m})\delta{m}-(ig_{ma} +\Gamma \mathrm{e}^{i\theta})\delta {a}-ig_{mb}m_{s}\delta{b},\nonumber\\ 
\delta\dot{b}& =- (i\omega_{b} + \gamma_{b}) \delta{b}-ig_{mb}m^{\ast}_{s}\delta{m}. \label{linerized-1}  
 \end{align}

\subsection{Exceptional point and $\mathcal{PT}$-symmetry}\label{SecIIC}
The  first-order linearized Heisenberg–Langevin equations in Eq.~(\ref{linerized-1}) can be written in the following matrix form:
\begin{equation}
 \dot{\mathbf{X}} = -i H_{\text{eff}}\,\mathbf{X},
\end{equation}
where $\mathbf{X} = (\delta a, \delta m, \delta b)^\mathsf{T}$ denotes the column vector of dynamical variables, and $H_{\text{eff}}$ corresponds to the effective non-Hermitian Hamiltonian governing the hybrid system,
\begin{equation}
H_{\text{eff}}=
{\begin{pmatrix}
\Delta_{a} -i\kappa_{a} & g_{ma} -i\Gamma\mathrm{e}^{i\theta} & 0 \\
g_{ma} -\mathrm{i\Gamma}\mathrm{e}^{i\theta}  & \Delta_{s} + i\kappa_{m}  &  G_{b} \\ 0 & G^*_{b} & \omega_{b} - i\gamma_{b} 
\end{pmatrix}},\label{H_eff}
\end{equation}
where $G_{b} = g_{mb}m_{s}$ is the effective magnon-phonon (magnomechanical) coupling coefficient. This hybrid three-mode system with magnon-photon and magnon-phonon couplings yields three eigenvalues. The eigenvalues $\lambda$ are determined by the characteristic equation:
\begin{equation}
\lambda^3 + r\lambda^2 + s\lambda + t=0,\label{cubic}
\end{equation}
where $r=-[\Delta_{a} -i\kappa_{a} +\Delta_{s} +i\kappa_{m} +\omega_{b} -i\gamma_{b}]$, $s=(\Delta_{a} -i\kappa_{a})(\Delta_{s} +i\kappa_{m} +\omega_{b} -i\gamma_{b}) + (\Delta_{s} +i\kappa_{m})(\omega_{b} -i\gamma_{b})-|G_{b}|^2 - (g_{ma} -i\Gamma\mathrm{e}^{i\theta})^2$, $t=-(\Delta_{a}-i\kappa_{a})(\Delta_{s} +i\kappa_{m})(\omega_{b} -i\gamma_{b}) + |G_{b}|^2(\Delta_{a} -i\kappa_{a}) +(\omega_{b} -i\gamma_{b})(g_{ma} -i\Gamma\mathrm{e}^{i\theta})^2$. 

Based on the effective Hamiltonian in Eq.~(\ref{H_eff}), the $\mathcal{PT}$-symmetric phases of the system were rigorously analyzed in Ref.~\cite{fahad2025} (Fig.~2). In this work, we investigate the GHS behavior that emerges across these $\mathcal{PT}$-symmetric regimes and at the $\mathrm{EP_3}$.

\subsection{Optical Susceptibility}
To evaluate the optical susceptibility of the coupled system, we solve Eq.~(\ref{linerized-1}) by introducing the slowly varying operators for the linear fluctuation terms: $\delta a \rightarrow \delta a\mathrm{e}^{-i\Delta_{a}t}$, $\delta m \rightarrow \delta m\mathrm{e}^{-i\Delta_{s}t}$, and $\delta b \rightarrow \delta b\mathrm{e}^{-i\omega_{b}t}$. We then apply the ansatz $\delta O = \delta O_{1}\mathrm{e}^{-ixt} + \delta O_{2}\mathrm{e}^{ixt}$ with $O= (a, m, b)$ and the effective detuning $x=\Delta_{p}-\omega_{b}$. This approach yields an explicit expression for the amplitude $\delta a_{1}$ of the first-order sideband in the non-Hermitian CMM system for a weak probe field:
\begin{equation}
\delta{a}_{1} = \frac{E_{p}}{(\kappa_{a}-ix) + \frac{(\gamma_{b}-ix)(ig_{ma} + \Gamma\mathrm{e}^{i\theta})^2}{(\kappa_{m}+ix)(\gamma_{b} -ix) - |G_{b}|^2}}. \label{amplitude} 
\end{equation}
The contribution of $\delta a_{2}$ is negligible, as it stems from a four-wave mixing process at frequency $\omega_{p} - 2\omega_{0}$ between the weak probe and the driving fields. The output field $E_{T}$ of the weak probe field defines the optical susceptibility $\chi$ via the relation~\cite{waseem_Goos_2024,MUNIR_2025,Chen_2023,Li_2016Transparency}
\begin{equation}
\chi \equiv E_T = \kappa_{a} \delta a_{1} / E_{p},\label{optical sus}
\end{equation}
where $\chi$ is a complex quantity, expressed in terms of its quadrature components as $\chi = \chi_{r} + i\chi_{i}$. 
These components are measured via homodyne detection~\cite{walls1994quantum}. The real ($\chi_{r}$) and imaginary ($\chi_{i}$) components of the $\chi$ describe the absorption and dispersion spectra of the probe field, respectively.
\subsection{Goos-Hänchen shift}
The TE-polarized probe field $E_{p}$  reflects from mirror $M_{1}$, resulting in a lateral displacement (GHS) $S_{r}$ along the $z$-axis. We employ the stationary phase method to quantify this displacement. Within this framework, the well-collimated probe field—with sufficiently narrow angular divergence—can be approximated as a plane wave. The GHS of the reflected probe field is then expressed as~\cite{Artmann-1948, Li-2003PRL}
\begin{equation}
S_{r} = -\frac{\lambda_{p}}{2\pi} \frac{d\phi_{r}}{d\theta_{i}},
\end{equation}
where $\lambda_{p}$ is the incident probe field wavelength, $\phi_{r}$ is the phase of the TE-polarized reflection coefficient $r(k_{z}, \omega_{p})$, $k_{z} = (2\pi / \lambda_{p}) \sin{\theta_{i}}$ is the $z$-component of the wavenumber, and $\theta_{i}$ is the probe field incidence angle. The GHS takes the explicit form~\cite{Wang:05, Wang_2008}
\begin{equation}
\begin{split}
S_{r} = -\frac{\lambda_{p}}{2\pi} \frac{1}{|r(k_{z},\omega_{p})|^2}
\left\{
\operatorname{Re}[r(k_{z},\omega_{p})] \frac{d \operatorname{Im}[r(k_{z}, \omega_{p})]}{d\theta_{i}}\right. \\
\left. - \operatorname{Im}[r(k_{z}, \omega_{p})] \frac{d \operatorname{Re}[r(k_{z}, \omega_{p})]}{d\theta_{i}}
\right\}.
\label{GHS} 
\end{split}
\end{equation}
The reflection coefficient $r(k_{z}, \omega_{p})$ used in Eq.~(\ref{GHS}) is obtained using the standard transfer-matrix method for the $j$-th layer system~\cite{Wang_2008}
\begin{equation}
M_{j}(k_{z},\omega_{p},d_{j}) =
\begin{pmatrix}
\cos[k_{x}^{j} d_j] & ik/ k_{x}^{j} \sin[k_{x}^{j} d_{j}] \\
i  k_{x}^{j}/k \sin[k_{x}^{j} d_{j}] & \cos[k_{x}^{j} d_{j}]
\end{pmatrix}, \label{transfer matrix}
\end{equation}
where $k_{x}^{j} = k \sqrt{\varepsilon_{j} - \sin^2{\theta_{i}}}$ is the $x$-component of the wavenumber of the probe field in the $j$-th layer. Here, $k = \omega_{p}/c$  is the wavenumber in vacuum, where $c$ is the speed of light. For each layer ($j\equiv1,2,3$), $\varepsilon_{j}$ and $d_{j}$ denote the permittivity and thickness, respectively. The effective intracavity permittivity $\varepsilon_{2} = 1 + \chi$ emerges from the nonlinear optical susceptibility $\chi$ [Eq.~\ref{optical sus}]. The total transfer matrix for the effective three-layer system is expressed as~\cite{Wang_2008} 
\begin{widetext}
\begin{equation}   
Q(k_{z}, \omega_{p}) = M_{1}(k_{z}, \omega_{p}, d_{1}) M_{2}(k_{z}, \omega_{p}, d_{2}) M_{1}(k_{z}, \omega_{p}, d_{1}) =\begin{pmatrix}
Q_{11} & Q_{12} \\
Q_{21} & Q_{22}
\end{pmatrix},
\end{equation}
\end{widetext}
where $M_{j}(k_{z}, \omega_{p}, d_{j})$ depends on the parameters of the corresponding layer. The reflection coefficient $r(k_{z}, \omega_{p})$ is given by
\begin{equation}
r(k_{z}, \omega_{p}) = \frac{q_{0}(Q_{22} - Q_{11}) - (q_{0}^2 Q_{12} - Q_{21})}{q_{0}(Q_{22} + Q_{11}) - (q_{0}^2 Q_{12} + Q_{21})}, \label{2x2}
\end{equation}
where $q_{0} = \sqrt{\varepsilon_{0} - \sin^2{\theta_{i}}}$, and $Q_{ij}$ (with $i, j = 1, 2$) denote the elements of the transfer matrix $Q(k_{z}, \omega_{p})$.
\section{Results and Discussion}
\begin{figure}[t]
\centering
\includegraphics[width=\linewidth]{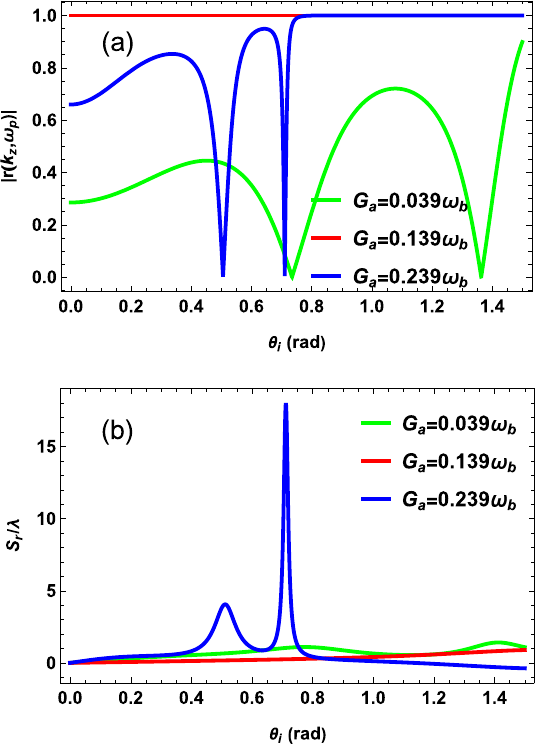}
\caption{(a) Absolute value of the reflection coefficient $|r(k_{z}, \omega_{p})|$ and (b) the normalized GHS $S_{r}/\lambda$ versus incident angle $\theta_{i}$ for three effective magnon-photon coupling regimes: (i) $G_{a} = 0.039\,\omega_{b}$ (green, broken $\mathcal{PT}$ phase), (ii) $0.139\,\omega_{b}$ (red, third-order exceptional point $\mathrm{EP}_3$), and (iii) $0.239\,\omega_{b}$ (blue, unbroken $\mathcal{PT}$ phase) at the resonance condition ($x=0$). Fixed parameters: $\omega_{a}/2\pi = 13.2~\mathrm{GHz}$, $\kappa_{a}/2\pi= 2.1~\mathrm{MHz}$, $\gamma_{b}/2\pi= 150~\mathrm{Hz}$, $\kappa_{m}=\kappa_{a} + \gamma_{b}$, $\omega_{b}/2\pi =15.101~\mathrm{MHz}$, $G_{b}/2\pi=0.001~\mathrm{MHz}$, $\epsilon_{0}=1$, $\epsilon_{1} =\epsilon_{3} = 2.2$, $d_{1} = 4~ \mathrm{mm}$, and $d_{2} = 45~\mathrm{mm}$.}
\label{fig2}
\end{figure}
\begin{figure}
\begin{center}
\includegraphics[width=\linewidth]{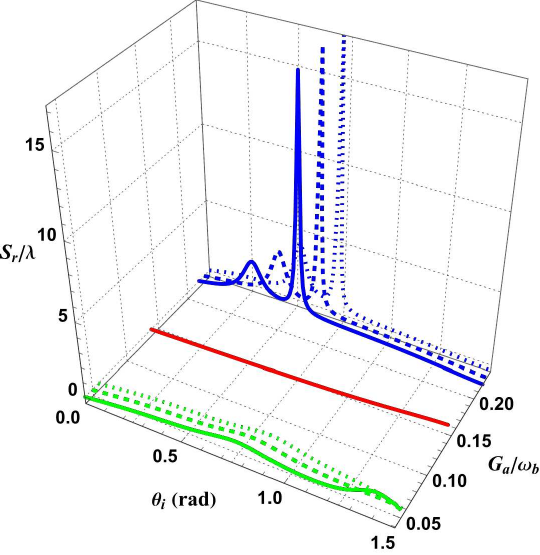}
\caption{Normalized GHS $S_r/\lambda$ as a function of the incident angle $\theta_i$, parametrized by the effective magnon-photon coupling strength $G_a$. The curves are shown for the three dynamical phases: broken $\mathcal{PT}$ phase, $\mathrm{EP}_3$, and the unbroken $\mathcal{PT}$ phase, respectively, at resonance ($x=0$). The effective coupling strengths are: $G_a/\omega_b = 0.039$, $0.049$, $0.058$ (green solid/dashed/dotted for the broken $\mathcal{PT}$ phase); $0.139$ (red solid at the $\mathrm{EP}_3$); and $0.22$, $0.229$, $0.239$ (blue solid/dashed/dotted for the unbroken $\mathcal{PT}$ phase). Fixed parameters are: $\omega_{a}/2\pi = 13.2~\mathrm{GHz}$, $\omega_{b}/2\pi =15.101~\mathrm{MHz}$, $\kappa_{a}/2\pi= 2.1~\mathrm{MHz}$, $\gamma_{b}/2\pi= 150~\mathrm{Hz}$, $\kappa_{m}=\kappa_{a} + \gamma_{b}$, $G_{b}/2\pi=0.001~\mathrm{MHz}$, $\epsilon_{0}=1$, $\epsilon_{1} =\epsilon_{3} = 2.2$, $d_{1} = 4~\mathrm{mm}$, and $d_{2} = 45~\mathrm{mm}$.}
\label{fig3}
\end{center}
\end{figure}
\begin{figure*}
\begin{center}
\includegraphics[width=5.5cm]{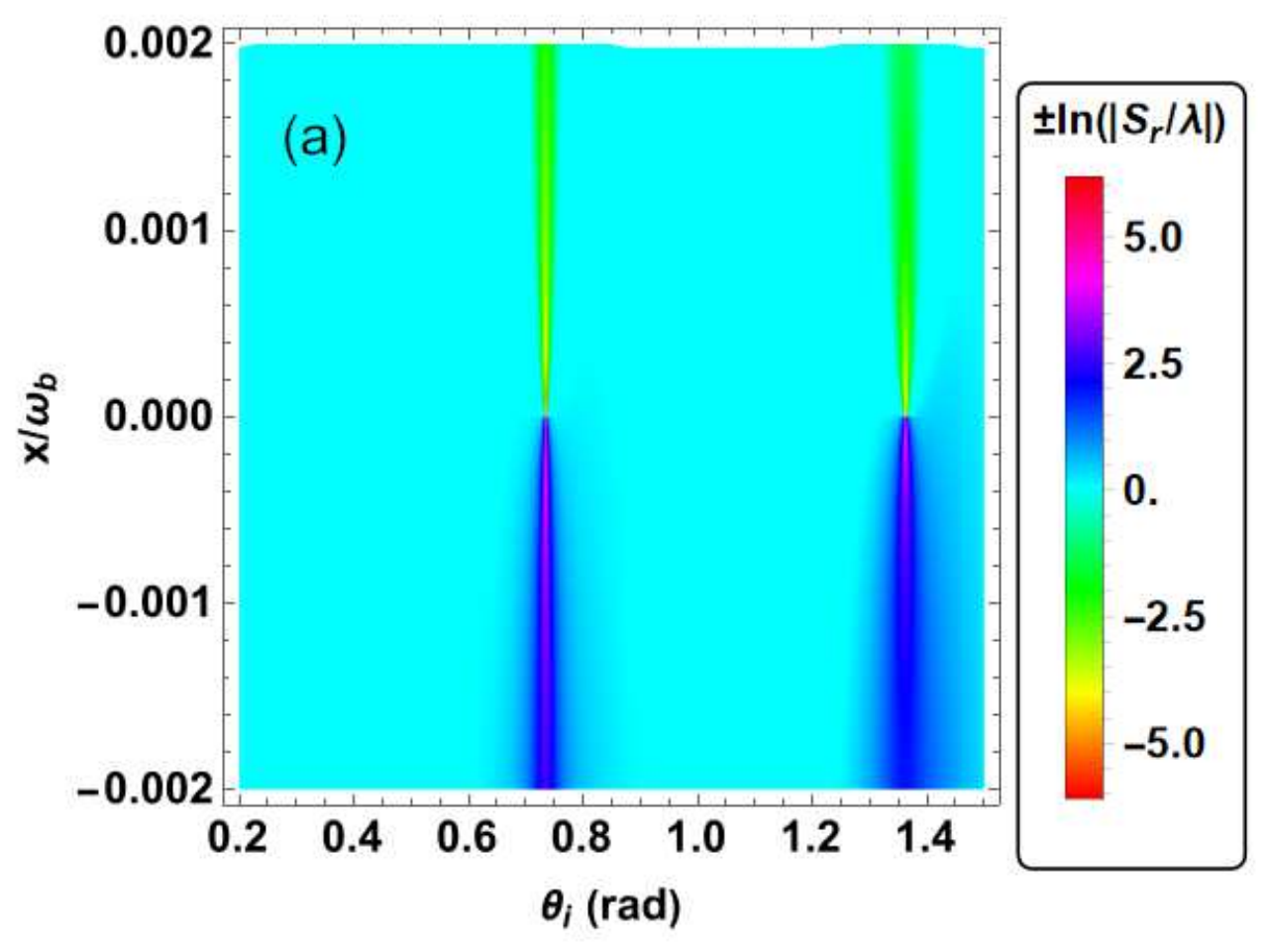}
\includegraphics[width=5.5cm]{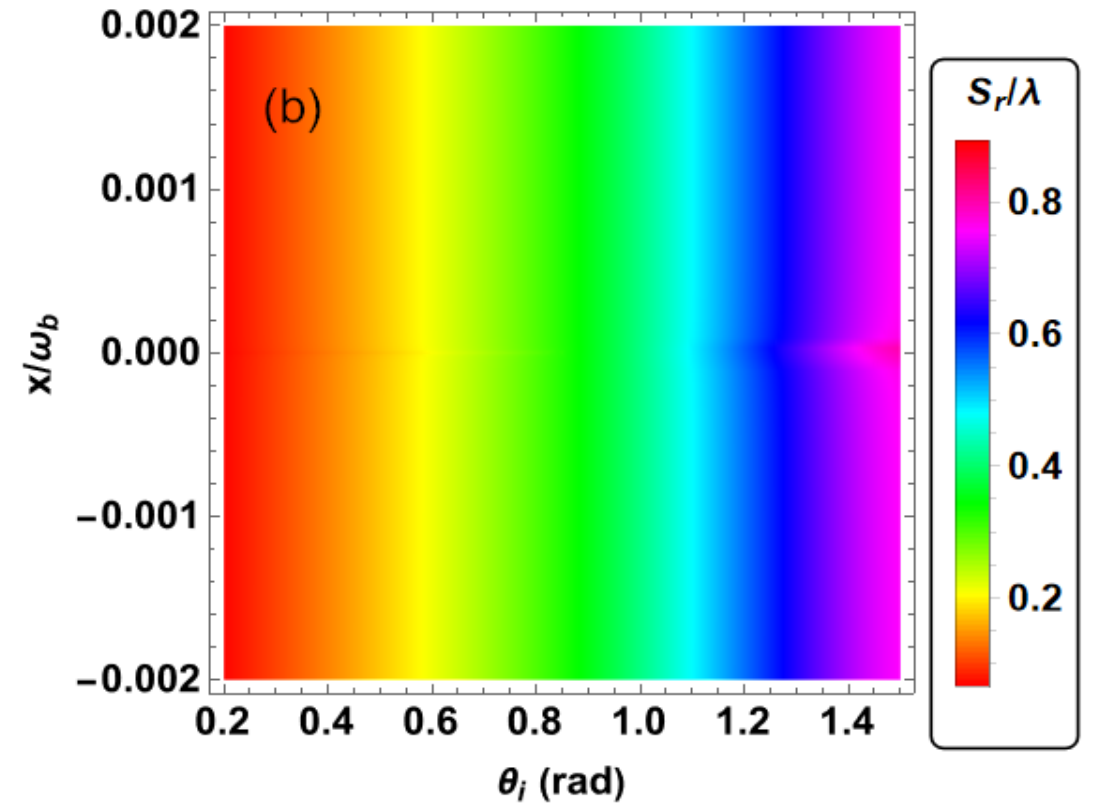}
\includegraphics[width=5.5cm]{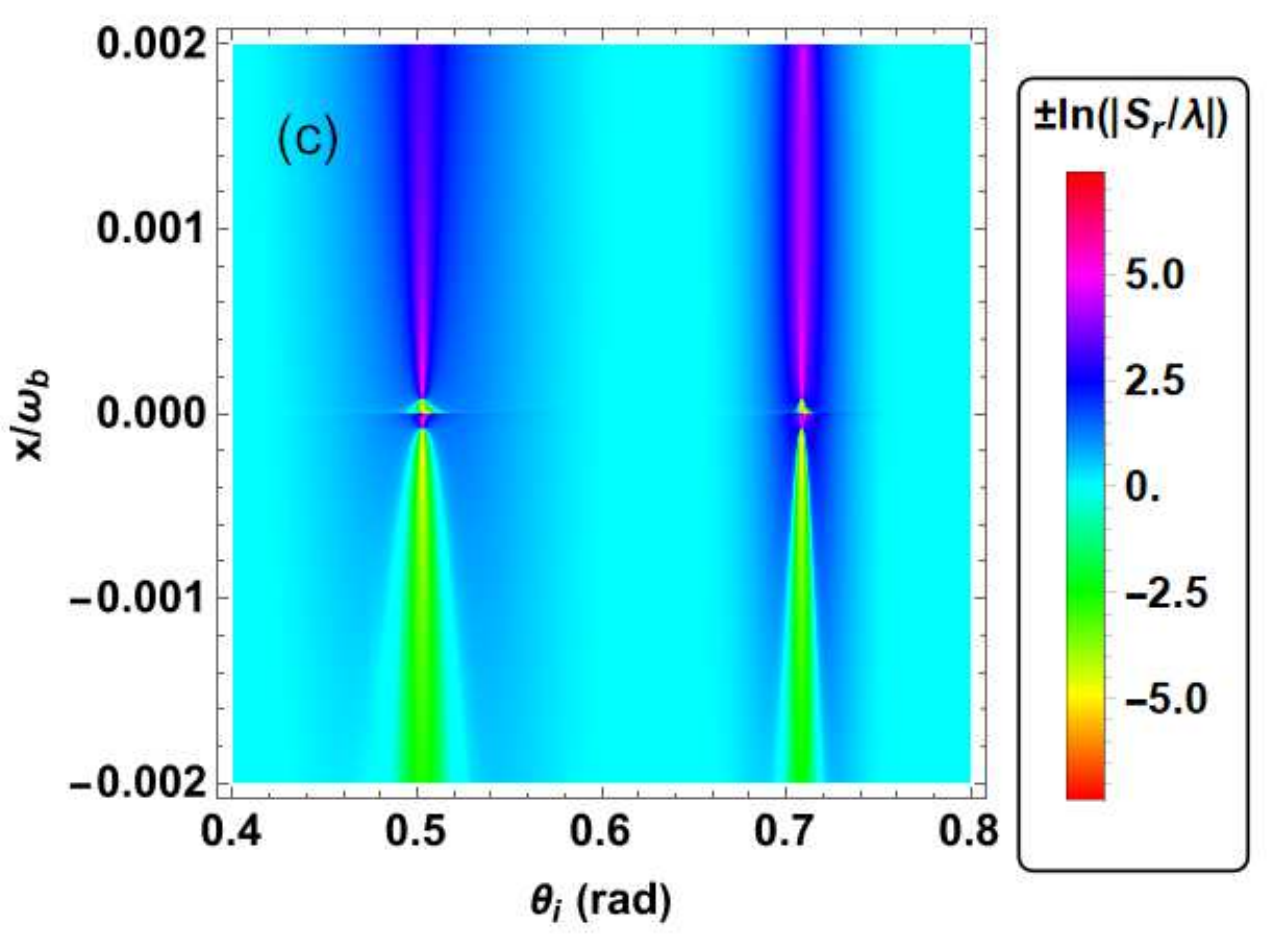}
\caption{Contour plots of normalized GHS $S_{r}/\lambda$ versus incident angle $\theta_{i}$ and normalized effective detuning $x/\omega_{b}$ for: (a) broken $\mathcal{PT}$ phase ($G_{a} = 0.039\,\omega_{b}$), (b) $\mathrm{EP}_3$ ($G_{a} = 0.139\,\omega_{b}$), and (c) unbroken $\mathcal{PT}$ phase ($G_{a} = 0.239\,\omega_{b}$). Fixed parameters: $\omega_{a}/2\pi = 13.2~\mathrm{GHz}$, $\omega_{b}/2\pi = 15.101~\mathrm{MHz}$, $\kappa_{a}/2\pi = 2.1~\mathrm{MHz}$, $\gamma_{b}/2\pi = 150~\mathrm{Hz}$, $\kappa_{m} = \kappa_{a} + \gamma_{b}$, $G_{b}/2\pi = 0.001~\mathrm{MHz}$, $\epsilon_{0}=1$, $\epsilon_{1} =\epsilon_{3} = 2.2$, $d_{1} = 4~\mathrm{mm}$, and $ d_{2} = 45~\mathrm{mm}$.}
\label{fig4}
\end{center}
\end{figure*}
\begin{figure*}
\centering
\includegraphics[width=\linewidth]{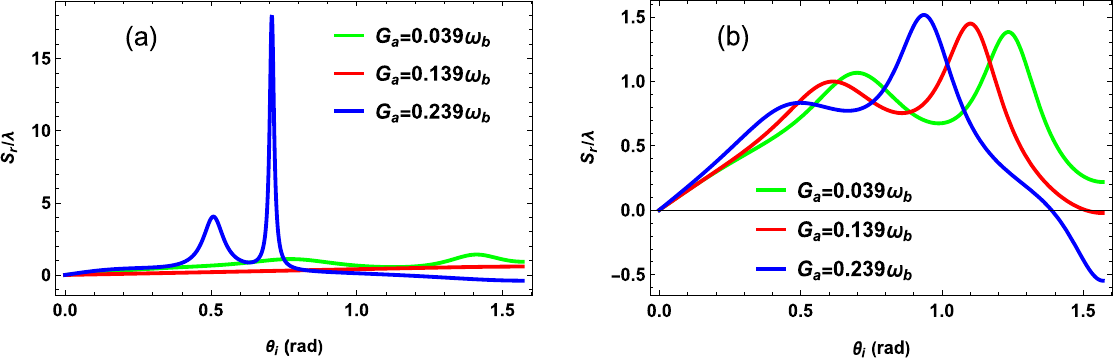}
\caption{Normalized GHS $S_{r}/\lambda$ versus incident angle $\theta_{i}$ for two cases: (a) $G_{b} =0$ and (b) $G_{b}/2\pi = 0.05~\mathrm{MHz}$ at resonance ($x=0$). Fixed parameters: $G_{a}= 0.039\,\omega_{b}$ (broken $\mathcal{PT}$ phase, green), $G_{a}=0.139\,\omega_{b}$ ($\mathrm{EP}_2$, red), $G_{a}=0.239\,\omega_{b}$ (unbroken $\mathcal{PT}$ phase, blue), $\omega_{a}/2\pi = 13.2~\mathrm{GHz}$, $\kappa_{a}/2\pi= 2.1~\mathrm{MHz}$, $\gamma_{b}/2\pi= 150~\mathrm{Hz}$, $\kappa_{m}=\kappa_{a} + \gamma_{b}$, $\omega_{b}/2\pi =15.101~\mathrm{MHz}$, $\epsilon_{0}=1$, $\epsilon_{1} =\epsilon_{3} = 2.2$, $d_{1} = 4~\mathrm{mm}$, and $ d_{2} = 45~\mathrm{mm}$.}
\label{fig5}
\end{figure*}
In this section, we present numerical results using experimentally feasible parameters~\cite{Harder_PRL, Zhang_2016_cavity}: cavity frequency $\omega_a / 2 \pi=13.2~\mathrm{GHz}$, mechanical frequency $\omega_b / 2 \pi=15.101~\mathrm{MHz}$, cavity decay rate $\kappa_a / 2 \pi=2.1~\mathrm{MHz}$, mechanical damping rate $\gamma_b / 2 \pi=150~\mathrm{Hz}$, magnon gain rate $\kappa_m=\kappa_a+\gamma_b$, and effective magnomechanical coupling coefficient $G_b / 2 \pi=0.001~\mathrm{MHz}$. The YIG sphere (diameter $D=250~\mu \mathrm{m}$, spin density $\rho=4.22 \times 10^{27} \mathrm{~m}^{-3}$, gyromagnetic ratio $\gamma / 2 \pi=28~\mathrm{GHz} / \mathrm{T}$) is driven by a magnetic field $B_0 \leq 0.5~\mathrm{mT}$, yielding $G_b / 2 \pi \leq 1.5~ \mathrm{MHz}$~\cite{Lu_2021_Ep}. For the GHS investigation, the dielectric constants are set to $\epsilon_0=1, \epsilon_1=\epsilon_3=2.2$, with layer thicknesses $d_1=4~\mathrm{mm}$ and $d_2=45~\mathrm{mm}$~\cite{Li_2020_phase-control, Zhang_2016_cavity}. The system exhibits $\mathcal{PT}$-symmetry exclusively when the angle between the traveling field and the cavity's $x$-axis is $\pi/2$.

Under balanced gain and loss conditions ($\kappa_m = \kappa_a + \gamma_b$) and with a finite effective magnomechanical coupling $G_b$, the eigenvalue spectrum of the non-Hermitian CMM system exhibits a $\mathcal{PT}$-symmetry phase transition, as described by Eq.~(\ref{H_eff}) and detailed in Ref.~\cite{fahad2025} (Fig.~2). This transition is tuned by the effective magnon–photon coupling strength $G_a$. For $G_a / \omega_b < 0.139$, the system resides in the broken $\mathcal{PT}$-symmetric phase, characterized by complex eigenvalues. At the critical coupling $G_a /\omega_b = 0.139$, an $\mathrm{EP}_3$ emerges, where both eigenvalues and eigenvectors coalesce. When $G_a / \omega_b > 0.139$, the eigenvalue spectrum becomes entirely real, indicating the unbroken $\mathcal{PT}$-symmetric phase. Thus, the numerical analysis of the GHS is systematically delineated across three regimes: (i) broken $\mathcal{PT}$-symmetric phase, (ii) $\mathrm{EP}_3$, and (iii) unbroken $\mathcal{PT}$-symmetric phase.

Equation~(\ref{GHS}) shows that the normalized GHS $S_{r}/\lambda$, depends on the reflection coefficient $|r(k_{z}, \omega_{p})|$ of the incident TE-polarized probe field. Figures~\ref{fig2}(a) and (b) display both $|r(k_{z}, \omega_{p})|$ and $S_{r}/\lambda$ as functions of the probe field incident angle $\theta_{i}$ for broken $\mathcal{PT}$ phase (green), $\mathrm{EP}_3$ (red), and unbroken $\mathcal{PT}$ phase (blue). Figure~\ref{fig2}(a) shows distinct reflection dips at specific incident angles of the probe field, corresponding to resonance conditions in both broken and unbroken $\mathcal{PT}$ phases. These resonant features indicate the existence of lateral shifts in both phases. The GHS peaks in broken and unbroken $\mathcal{PT}$ phases coincide with the reflection resonances of the probe field [Fig.~\ref{fig2}(b)]. In the unbroken $\mathcal{PT}$ phase, the GHS is significantly enhanced, exhibiting a large positive shift. In contrast, the broken $\mathcal{PT}$ phase yields much smaller positive shifts across the entire angular range. This difference originates from the strong effective magnon-photon coupling in the unbroken $\mathcal{PT}$ phase, which yields sharper reflection resonances and steeper phase gradients, thereby amplifying the GHS. The observed GHS enhancement agrees with previous reports for atomic-ensemble-based $\mathcal{PT}$-symmetric cavities~\cite{Zia_2015_PRA}. On the other hand, the weaker effective coupling in the broken phase leads to broader spectral features and reduced phase dispersion, thus suppressing the shift [Fig.~\ref{fig2}(b)]. At the $\mathrm{EP}_3$, the coalescence of eigenvalues and their corresponding eigenvectors flattens the reflection phase ($|r(k_{z}, \omega_{p})|\approx1$), eliminating phase dispersion and causing the GHS to approach zero.

\begin{figure*}
\begin{center}
\includegraphics[width=5.8cm]{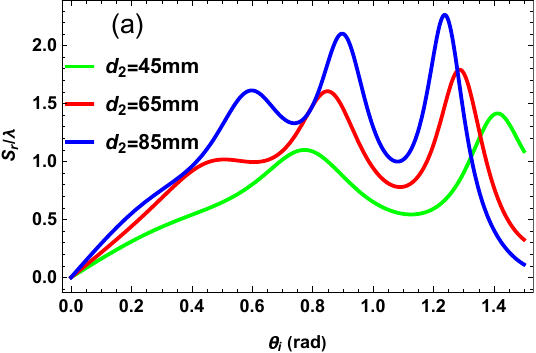}
\includegraphics[width=5.8cm]{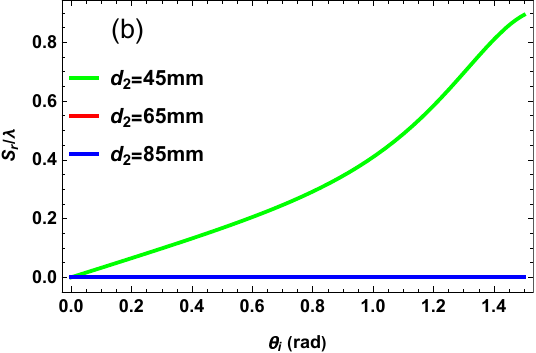}
\includegraphics[width=5.8cm]{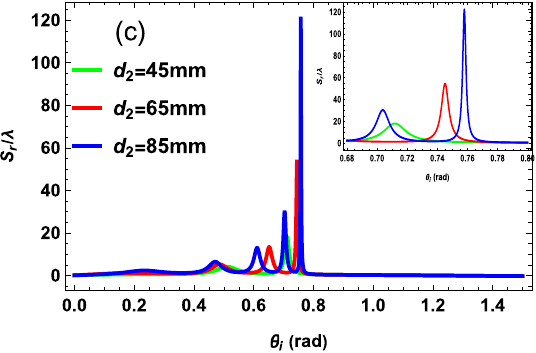}
\caption{Normalized GHS $S_{r}/\lambda$ as a function of incident angle $\theta_{i}$ for (a) broken $\mathcal{PT}$ phase ($G_{a} = 0.039\,\omega_{b}$), (b) $\mathrm{EP}_3$ ($G_{a} = 0.139\,\omega_{b}$), and (c) unbroken $\mathcal{PT}$ phase ($G_{a} = 0.239\,\omega_{b}$), respectively at three different intracavity medium lengths. Green, red, and blue curves show the GHS at $d_{2}=45$, $65$, and $85~\mathrm{mm}$, respectively, at the resonance condition ($x=0$). The insets in Fig.~\ref{fig6}(c) provide magnified views of $S_{r}/\lambda$ as a function of $\theta_{i}$ in $\mathcal{PT}$-symmetric phase at different intracavity medium lengths. Fixed parameters are $\kappa_{a}/2\pi=2.1~\mathrm{MHz}$, $\gamma_{b}/2\pi= 150~\mathrm{Hz}$, $\kappa_{m}=\kappa_{a} + \gamma_{b}$, $\omega_{b}/2\pi =15.101~\mathrm{MHz}$, $ G_{b}/2\pi=0.001~\mathrm{MHz}$, $\omega_{a}/2\pi = 13.2~\mathrm{GHz}$, $d_{1} = 4~\mathrm{mm}$, $\epsilon_{0}=1$, and $\epsilon_{1} = \epsilon_{3} = 2.2$.}
\label{fig6}
\end{center}
\end{figure*}
\begin{figure}
\centering
\includegraphics[width=\linewidth]{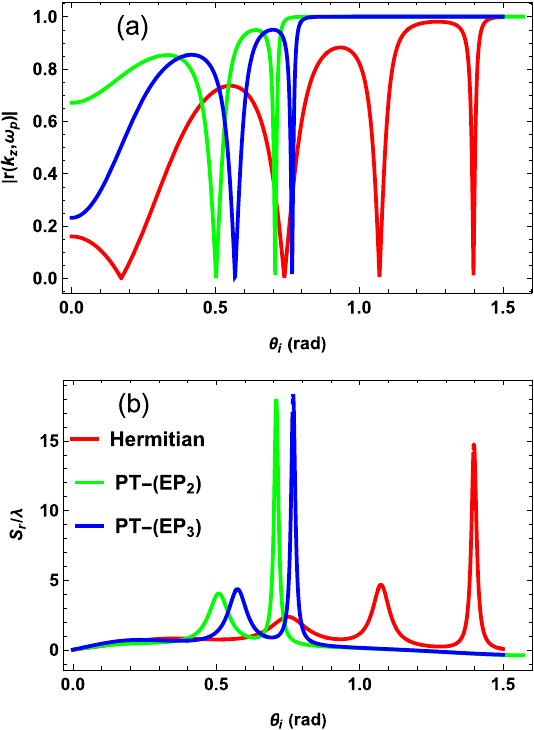}
\caption{(a) Reflection coefficient $|r(k_{z}, \omega_{p})|$ and (b) the normalized GHS $S_{r}/\lambda$ versus incident angle $\theta_{i}$ for Hermitian case ($G_{b}/2\pi=0.005~\mathrm{MHz}$, red curves, $\kappa_{m}/2\pi= 0.1~\mathrm{MHz}$), and $\mathcal{PT}$-symmetric system ($G_{b}=0$, $\mathrm{EP}_2$, green curves, and $G_{b}/2\pi=0.005~\mathrm{MHz}$, $\mathrm{EP}_3$, blue curves, $\kappa_{m} = \kappa_{a} + \gamma_{b}$) at the resonance condition ($x=0$). Fixed parameters: $G_{a}=0.239\,\omega_{b}$, $\omega_{a}/2\pi = 13.2~\mathrm{GHz}$, $\kappa_{a}/2\pi= 2.1~\mathrm{MHz}$, $\gamma_{b}/2\pi= 150~\mathrm{Hz}$, $\omega_{b}/2\pi =15.101~\mathrm{MHz}$, $\epsilon_{0}=1$, $\epsilon_{1} =\epsilon_{3} = 2.2$, $d_{1} = 4~\mathrm{mm}$, and $d_{2} = 45~\mathrm{mm}$.}
\label{fig7}
\end{figure}

In Figs.~\ref{fig2}(a) and (b), we analyze representative values of the effective magnon-photon coupling strength $G_a$ corresponding to the broken $\mathcal{PT}$ phase, $\mathrm{EP}_3$, and the unbroken $\mathcal{PT}$ phase. To generalize these results, we explore a broader range of $G_a$ within each regime and observe consistent qualitative behavior: the GHS is significantly enhanced in the unbroken $\mathcal{PT}$ phase, strongly suppressed in the broken phase, and approaches zero at the $\mathrm{EP}_3$ [Fig.~\ref{fig3}]. The angular dependence and magnitude of the GHS in each regime are governed by $G_a $, illustrating tunable control over this lateral shift. This reproducibility across the parameter space highlights the robustness of the results and emphasizes the central role of $\mathcal{PT}$-symmetry in modulating the GHS.

Next, we analyze the contour plots of the GHS $S_{r}/\lambda$ as a function of both the incident angle $\theta_{i}$ and the normalized effective detuning $x/\omega_{b}$. Figures~\ref{fig4}(a-c) show the $S_{r}/\lambda$ versus $\theta_{i}$ and  $x/\omega_{b}$ in the broken $\mathcal{PT}$ phase, $\mathrm{EP}_3$, and unbroken $\mathcal{PT}$ phase, respectively. For the broken and unbroken $\mathcal{PT}$ phases, we employed a minimal step size (for both variables) to generate smooth, high-resolution plots [Figs.~\ref{fig4}(a) and (b)]. Consequently, some peaks and dips appear sharply defined with large amplitudes. In the broken $\mathcal{PT}$ phase [Fig.~\ref{fig4}(a)], weak effective magnon–photon coupling produces complex eigenvalues, leading to asymmetric GHS. This results in a positive lateral shift for negative cavity detunings and a negative shift for positive detunings. At the $\mathrm{EP}_3$ [Fig.~\ref{fig4}(b)], the coalescence of eigenvalues and eigenvectors flattens the reflection phase, thereby suppressing the GHS. In contrast, the unbroken $\mathcal{PT}$ phase [Fig.~\ref{fig4}(c)] exhibits real eigenvalues and a sharp, symmetric GHS enabled by strong, coherent effective magnon-photon coupling, where the shift manifests as symmetric bands of large positive and negative lateral displacements centered at resonance ($x=0$).

Figures~\ref{fig5}(a) and (b) illustrate the normalized GHS $S_r/\lambda$ as a function of the probe field's incident angle $\theta_i$ for different effective magnomechanical coupling coefficients $G_b$. For $G_b = 0$, the system exhibits a distinct phase transition in the GHS [Fig.~\ref{fig5}(a)], consistent with the $\mathcal{PT}$-symmetric eigenvalue spectrum characteristic of an $\mathrm{EP}_2$ system~\cite{fahad2025} (Fig.~3). In contrast, at $G_b = 2\pi \times 0.05~\mathrm{MHz}$, the effective magnomechanical interaction significantly suppresses the overall GHS, eliminating the phase transition while concurrently enhancing the magnitude of the negative shift at higher incident angles [Fig.~\ref{fig5}(b)]. This suppression is attributed to increased absorption of the probe field at higher $G_b$~\cite{waseem_Goos_2024}. Furthermore, the transition of the GHS from positive to negative is associated with the group index of the overall cavity system~\cite{Zia_2010_Coherent, waseem_Goos_2024}.

The GHS $S_{r}/ \lambda$ depends crucially on the cavity geometry, particularly the total thickness $L = 2d_1 + d_2$, which necessitates precise dimensional control. Figures~\ref{fig6}(a-c) show $S_r/ \lambda$ versus the incident angle $\theta_i$ at resonance ($x=0$) for intracavity lengths of $d_{2} = 45~\mathrm{mm}$, $65~\mathrm{mm}$, and $85~\mathrm{mm}$.
In the broken $\mathcal{PT}$ phase, both the GHS magnitude and the number of resonant peaks increase with $d_2$ [Fig.~\ref{fig6}(a)]. At the $\mathrm{EP}_3$ [Fig.~\ref{fig6}(b)], the GHS is suppressed and approaches zero as $d_2$ increases. While the unbroken $\mathcal{PT}$ phase exhibits enhanced GHS magnitudes and a stronger dependence on $d_2$ than its broken phase [Fig.~\ref{fig6}(c)]. This intracavity length-dependent behavior in non-Hermitian cavity magnomechanics is analogous to the earlier observations in atomic systems~\cite{Zia_2010_Coherent} and Hermitian cavity magnomechanics~\cite{waseem_Goos_2024}, confirming the importance of cavity dimensions for controlling the GHS.

Finally, we compare the reflection coefficient magnitude $|r(k_{z}, \omega_{p})|$ and the normalized GHS $S_{r}/\lambda$ for both Hermitian and $\mathcal{PT}$-symmetric systems containing $\mathrm{EP}_2$ and $\mathrm{EP}_3$. The Hermitian case, corresponding to $\Gamma = 0$ in Eq.~(\ref{simplified-H}), reproduces the optical susceptibility described in Ref.~\cite{waseem_Goos_2024}. Figures~\ref{fig7}(a) and (b), which show $|r(k_{z}, \omega_{p})|$ and $S_r/\lambda$ respectively, indicate that the $\mathcal{PT}$-symmetric system—encompassing both $\mathrm{EP}_2$ (green) and $\mathrm{EP}_3$ (blue)—exhibits a pronounced enhancement of the GHS relative to the Hermitian case (red). This enhancement arises from the gain–loss balance, which increases the reflection-phase sensitivity. As a result, the $\mathcal{PT}$-symmetric configuration provides greater control over the amplitude and phase of the reflected probe field, producing larger GHS [Fig.~\ref{fig7}(b)] than those achieved in Hermitian systems~\cite{waseem_Goos_2024}. Moreover, the shift associated with $\mathrm{EP}_3$ is comparatively larger than that at $\mathrm{EP}_2$, reflecting stronger phase sensitivity in higher-order $\mathrm{EP}$ systems. These results underscore the advantages of $\mathcal{PT}$-symmetric structures for applications demanding precise angular control and selective signal routing, outperforming conventional photonic systems.

\section{Conclusion}
In summary, we have investigated the GHS of a reflected probe field in a hybrid non-Hermitian CMM system. The system exhibits three distinct phases—(a) broken $\mathcal{PT}$ symmetry, (b) third-order exceptional point $\mathrm{EP}_3$, and (c) unbroken $\mathcal{PT}$ symmetry, which collectively establish a controlled platform for exploring phase-dependent lateral-shift phenomena.

Analysis at resonance ($x = 0$) revealed that the GHS magnitude was substantially enhanced in the unbroken $\mathcal{PT}$ phase compared to the broken $\mathcal{PT}$ phase, consistent with previous findings \cite{Zia_2015_PRA}. In contrast, at the $\mathrm{EP}_3$—characterized by the coalescence of eigenvalues and eigenvectors—the GHS demonstrated strong suppression. This qualitative behavior is consistent across a broad range of effective magnon-photon coupling, with the GHS remaining enhanced in the unbroken phase, suppressed in the broken phase, and vanishing at the $\mathrm{EP}_3$. Furthermore, we have shown that the effective magnomechanical coupling is a critical parameter: in its absence, the system exhibits an $\mathrm{EP}_2$, and the GHS undergoes a clear phase transition. However, when this effective magnomechanical coupling is increased beyond a parametric window, the phase transition disappears, and strong absorption emerges at resonance, leading to a decrease in the magnitude of the GHS. Using the intracavity length as an active tuning parameter, we further demonstrate that the GHS increases with intracavity length in both the broken and unbroken $\mathcal{PT}$ phases—though with markedly different magnitudes—while remaining suppressed at the $\mathrm{EP}_3$. Comparative analysis between $\mathcal{PT}$-symmetric and Hermitian configurations demonstrated significant enhancement of the GHS in the PT-symmetric system relative to the Hermitian case~\cite{waseem_Goos_2024}. The pronounced enhancement of the GHS in the $\mathcal{PT}$-symmetric systems, resulting from precisely engineered gain and loss, underscores a distinct advantage over conventional Hermitian systems for applications necessitating substantial lateral shifts, such as quantum switching and high-precision microwave sensing. Our analysis provides valuable insights into the GHS in non-Hermitian CMM systems and may serve as a foundation for the development of microwave devices based on GHS effects.

\section*{Acknowledgement}
We acknowledge the financial support from the NSFC under grant 
No. 12174346.\\
\\

\section*{references}

\bibliography{ref}

\end{document}